\documentclass[12pt]{article}
\usepackage{amsmath}
\usepackage{enumerate}
\usepackage{natbib}
\usepackage{url} 
\usepackage{graphicx,psfrag,epsf,color}
\usepackage{amsfonts}
\usepackage{booktabs,multirow} 
\usepackage{subfig}
\usepackage{amsthm}
\usepackage[colorlinks,linkcolor=red,anchorcolor=green,citecolor=blue]{hyperref}
\def\t{\top}
\def\cdn{{|}} 
\def\v{{\varepsilon}} 
\def\pr{{ \mathbb{P} }} 
\newcommand{\indep}{\rotatebox[origin=c]{90}{$\models$}} 
\newtheorem{thm}{Theorem}

\newtheorem{lem}{Lemma}
\newtheorem{rem}{Remark}
\newtheorem{assm}{Condition}

\newcommand{\blind}{0}
\begin{document}
	
	\def\spacingset#1{\renewcommand{\baselinestretch}%
		{#1}\small\normalsize} \spacingset{1}
	
	\date{}
	\if0\blind
	{
		\title{\bf Adaptive-to-model hybrid of tests for regressions}
		\author{Lingzhu LI$^{1}$, Xuehu Zhu$^{2}$ and Lixing ZHU$^{1,3}$\thanks{
				The authors gratefully acknowledge a grant from the University Grants Council of Hong Kong and a NSFC grant (NSFC11671042). }\hspace{.2cm}\\
			$^1$Department of Mathematics, Hong Kong Baptist University, Hong Kong\\
			$^2$School of Mathematics, Xi’an Jiaotong University, Xi’an, China\\
			$^3$School of Statistics, Beijing Normal University, Beijing, China}
		\maketitle
	} \fi
	
	\if1\blind
	{
		\bigskip
		\bigskip
		\bigskip
		\begin{center}
			{\LARGE\bf Adaptive-to-model hybrid test for regressions}
		\end{center}
		\medskip
	} \fi
	
	\bigskip
\begin{abstract}
		In model checking for regressions, nonparametric estimation-based tests usually have tractable limiting null distributions and are sensitive to oscillating alternative models, but suffer from the curse of dimensionality. In contrast, empirical process-based tests can, at the fastest possible rate, detect local alternatives distinct from the null model, but is less sensitive to oscillating alternative models and with intractable limiting null distributions.
		It has long been an issue on how to construct a test that can fully inherit the merits of these two types of tests and avoid the shortcomings. We in this paper propose a generic adaptive-to-model hybrid of  moment and conditional moment-based test to achieve this goal.
		Further, a significant feature of  the method is to make  nonparametric estimation-based tests,  under the alternatives,  also share the merits of existing empirical process-based tests.
		This  methodology can be readily applied to other kinds of data and  constructing other hybrids. As a by-product in sufficient dimension reduction field, the estimation of residual-related central subspace  is used to indicate the underlying models for model adaptation. A systematic study is devoted to showing when  alternative models can be indicated and when cannot. This estimation  is of its own interest and can be applied to the problems with other kinds of data. Numerical studies are conducted to verify the powerfulness of the proposed test.
\end{abstract}
	
\noindent%
\vfill
	
\newpage
\spacingset{1.5} 
\section{Introduction}\label{sec:introduction}

Suppose that $X \in R^p$ is the $p$-dimensional  explanatory vector and $Y \in R $  the scalar response variable.
Without further information from the data, the underlying regression model can be written as
\begin{eqnarray} \label{underlying model}
  Y = G(X,\v),
\end{eqnarray}
where $G(\cdot)$ is a general nonparametric smooth function and $\v$ is the error term independent of $X$.
Such a model is too flexible to be interpreted, therefore in practical applications, a parametric model is often preferred:
\begin{eqnarray} \label{general parametric null function}
  Y = g(X,\theta) + \v,
\end{eqnarray}
where $g(\cdot, \cdot)$ is a given function up to a parameter $\theta \in R^d $.
A natural issue is to check whether this model is an adequate depiction of the  relationship between $X$ and  $Y$, since a wrongly specified model will cause unreliable statistical inferences. In this paper, we focus on this homogeneous null model. For more general heterogeneous models, we will give a discussion in Section~\ref{sec: discussions}.

Many efforts have been devoted to this model checking problem since 1980's with a number of tests available. To demonstrate the necessity of revisiting this issue, we first give a brief review of the pros and cons of existing methodologies and what are the difficulties any existing test cannot well handle. In the literature, two classes of the most popularly used methods in this area are nonparametric estimation-based tests which are the so-called local smoothing tests.
Early references of local smoothing tests include \cite{hardle1993comparing}, \cite{zheng1996consistent}, \cite{fan1996consistent} and \cite{koul2004minimum}, among others. \cite{hart1997nonparametric} is a comprehensive reference for the early research in this field.
The most appealing feature of local smoothing tests is their sensitivity to high-frequency/oscillating  alternative models.
However, the typical drawback is the slow rates of convergence, which causes low power performance when the dimension is high.
For instance, the test statistic in \cite{zheng1996consistent}, multiplied by $nh^{p/2}$, converges to  its weak limit where $n$ is the sample size and $h\to 0$ is the bandwidth in the Nadaraya–Watson kernel estimation of the conditional moment $E(Y-g(X,\theta)\cdn X)$. The rate $nh^{p/2}$ can be much slower than $ n$ when the dimension $p$ is even moderate and then causes that they can  only detect the local alternatives distinct from the null at the rate slower than $n^{-1/2}h^{-p/4}$. Thus, the dimension of $X$ severely worsens the power performance (see \cite{guo2016model}).
To alleviate this difficulty,
 \cite{lavergne2012one} used the projection method that  can  detect the local alternatives at the rate of  order $n^{-1/2} h^{-1/4}$ in theory. Their test involves the approximation for high-dimensional integral, which is computationally expensive (see \cite{zhu1995test}).
Further, when the approximation  is not sufficiently good or the dimension of $X$ is high, the test cannot maintain the significance level well.
Under a dimension reduction structure, \cite{guo2016model} suggested an adaptive-to-model approach that can also reach the rate of order $n^{-1/2} h^{-1/4}$, without that computational issue. However,  both are still (actually this is the case for any local smoothing test in the literature) impossible to reach the fastest possible rate of order $1/\sqrt n$ to detect local alternatives.

Unlike local smoothing tests,  global smoothing  test statistics are typically the averages 
of weighted sums of residuals. They are called the global smoothing tests as  averaging  in the CUSUM nature is a global smoothing step.
The most significant advantage of  global smoothing tests is that they can detect local alternatives at the fastest possible rate of $n^{-1/2}$ and converge to its weak limits at this rate (or $1/n$ if the quadratic form is used).
However, the main shortcomings of  global smoothing methods are 1) the CUSUM structure often weakens their ability to detect oscillating alternatives; 2) the  intractability of the limiting null distributions (see \cite{stute1998bootstrap,dette2007new}) requires  the assistance from Monte Carlo approximation/resampling technique for critical value determination unless  the  dimension $p$ is $1$ \citep{stute1997nonparametric} or a directional test is used \citep{stute2002model}, which is time consuming.

Above expounds the well-known pros and cons of these two types of tests showing that no methodology in the literature can simultaneously inherit the respective advantages and avoid the shortcomings. In summary,  having a test that has the following three  features is desirable:
\begin{itemize}
\item[$f_1$)] at the rate of $1/\sqrt n$,  the constructed test statistic converges to  a tractable weak limit such that the critical value can be easily determined without the assistance of Monte Carlo approximation/resampling technique;
\item[$f_2$)]  the test can be sensitive to oscillating alternatives as local smoothing tests, but less influenced by dimensionality;
\item[$f_3$)]  more than the omnibus property, at the fastest possible rate of order $1/\sqrt n$, the test can  detect local alternatives distinct from the null model, and thus is more powerful than any existing local smoothing tests in theory.
\end{itemize}
   To the best of our knowledge, any single test in the literature cannot share all these appealing features. To arrive the above goals, the test must have both local and global smoothing component. But due to their different convergence rates,  any simple convex combination
   does not work for this purpose. We will see this clearly in Sections~\ref{sec: proposed test} and \ref{sec: asymptotics}.

In this paper, we attempt to attack this longstanding problem. To be specific, we  propose an adaptive-to-model hybrid of tests that is a combination of moment-based test and nonparametric conditional moment-based test component. This hybrid can automatically indicate the underlying model such that it can decide which component
works
and then fully inherit the merits of local and global smoothing tests described in the above features $f_1$)-$f_3$). 
Model adaptation is achieved through an indicative dimension of a residual-based central subspace for the underlying model in the sense of sufficient dimension reduction. Thus, under the null hypothesis, it derives a {\it zero-dimensional projection} such that the test becomes a simple moment-based test for critical value determination. Under the alternatives, the test automatically becomes a nonparametric conditional moment-based omnibus test. More interestingly and importantly, the special construction of the test  makes the conditional moment-based component  achieve the fastest possible rate of convergence.
This is the most significant contribution of the proposed method as any existing nonparametric estimation-based test is impossible to have such a feature.
The test is based on the simplest moment test and a typical nonparametric conditional moment test proposed by \cite{zheng1996consistent}. In effect, we can choose any nonparametric  estimated-based test in the hybrid. The asymptotic properties could be similar. We will have some more discussion in Section~\ref{sec: discussions}.

For model adaptation, we  consider a residual-based central subspace  with indicative dimension. This concept is slightly different from that in sufficient dimension reduction field \citep{li1991sliced, cook1998regression} because this dimension can indicate the underlying model such that  the proposed hybrid  can do so.
As a by-product, we will propose a target matrix and suggest a criterion to define
an estimator of the  indicative dimension. 
The relevant properties show that the estimated dimension
can adapt to
the underlying model even when  the local alternative models approximate to the null model at the optimal rate that is as close to $n^{-1/2}$ as possible in a certain sense.
It
improves existing results for model adaptation to the underlying model through the dimension indication at the rate of order slower than $n^{-1/2}h^{-1/4}$, see  \cite{guo2016model} even when a dimension reduction structure is assumed.
The desirable asymptotic properties of the proposed hybrid test are derived under the null, local and global alternative hypothesis.
The resulting test can be very different from the adaptive-to-model one in \cite{guo2016model} which cannot change its local smoothing nature and is  impossible to inherit the merits of global smoothing tests.

It is worthwhile to mention that this generic methodology is ready to apply to any pair of tests to obtain the hybrid. Also, it can be applied to other kinds of data such as measurement error data, panel data and functional data. We will have a brief discussion in Section~\ref{sec: discussions}. 

The materials in this paper are organized as follows.
In Section~\ref{sec: proposed test}, we present the  hypotheses and the test statistic construction.
A target matrix  and a criterion for estimating indicative dimension are suggested in Section \ref{sec: dimension estimation}.
The various rates under the null, local and global alternative hypothesis for dimension indication are systematically studied and the optimal indicative rate  is derived.
Section~\ref{sec: asymptotics} contains the asymptotic properties of the test statistic. 
Numerical performances of the test under different models are examined by various experiments in Section~\ref{sec:NumericalStudy}.
Section~\ref{sec: discussions} includes some discussions about the main limitations  and possible extension to heteroscedastic models.
The regularity conditions of the theorems are listed in the Appendix~\ref{appendix}. The proofs for the theorems and details of dimension estimation are presented in the supplementary materials\footnote{See Supplementary materials.}.

\section{The test statistic construction} \label{sec: proposed test}

\subsection{A brief review of sufficient dimension reduction}
Suppose there are two random column vectors $Z_1 \in R^{p_1}$ and $Z_2\in R^{p_2}$.
If there exists a matrix $B \in R^{p_1 \times q}$, such that, $Z_2 \indep Z_1 \cdn B^\t Z_1 $.
Then the column space $Span\{B\}$ of $B$ is called a sufficient dimension reduction subspace of $Z_2$ with respect to $Z_1$.
The intersection of all the dimension reduction subspaces is called the central subspace and denoted as $S_{Z_2\cdn Z_1}$.
The dimension of the central subspace is denoted as $dim(S_{Z_2 \cdn Z_1})$.
If $B$ is the central subspace, then its column dimension $q = dim(S_{Z_2 \cdn Z_1})$.
When the real working dimension $q$ is smaller than the original dimension $p_1$ of $Z_1$ and dimension reduction can be achieved in terms of identifying the matrix $B$ and its dimension $q$. See \cite{li1991sliced} and \cite{cook1998regression} for more details.
We will use these notations during the test construction next.

\subsection{The hypotheses}

Based on \eqref{underlying model} and \eqref{general parametric null function}, the null hypothesis we concern about is
\begin{eqnarray}\label{Hypo_0}
	H_0 : Y = g(X,\theta_0)+\v,\quad \text{for some }\theta_0\in \Theta,
\end{eqnarray}
where $\Theta$ is a subset of $R^d$ and $\v\indep X$ is the error term.
The alternative hypothesis is
\begin{eqnarray}\label{Hypo_1}
	H_1 : Y = G(X,\v) \neq g(X,\theta)+\v, \quad \text{for all }\theta\in \Theta.
\end{eqnarray}
Let $\theta^\ast = \arg\min_{\theta\in\Theta}E\{Y-g(X,\theta)\}^2$ and $\eta = Y-g(X,\theta^\ast)$ provided that the involved second order moments exist.
Under the conditions in  Appendix, 
the minimizer $\theta^\ast$ is identifiable and can be consistently estimated.
Under the null hypothesis, we have $\eta = \v$, $\theta^\ast =\theta_0$ and then $dim(S_{\eta\cdn X}) = 0$ by the independence between $\eta$ and $X$.
While under the alternative hypothesis, the residual $\eta = G(X,\v)-g(X,\theta^\ast)-\v = \Delta(X,\v)$, which leads to $dim(S_{\eta\cdn X}) > 0 $ since $\Delta(X,\v)$ is a non-constant function of $X$.
Thus,   $dim(S_{\eta\cdn X})$ is respectively equal to $0$ and larger than $0$ under the null and alternative hypothesis.
This inspires us to construct a test that can fully use the information provided by  this indicative dimension. In the following, we implement our idea.

\subsection{The hybrid of tests}

We in this subsection describe our idea in terms of using two simple tests as the components in the hybrid. Following description essentially provides a general framework that can be used to develop other hybrids.

Suppose $\{(x_1, y_1), \dots, (x_n,y_n)\}$ is an available sample, where $(x_i,y_i), i=1,\dots,n$ are independent and identically distributed.
For the sake of illustration, the unknown parameter $\theta$ is estimated by the least squares method.
Define $\hat\theta_n = \arg\min_{\theta\in \Theta} \sum_{j=1}^n [ y_j-g(x_j,\theta) ]^2  $ and $\hat \eta_j = y_j - g(x_j,\hat\theta_n)$.

 To achieve the optimal rate of convergence and the  tractability of the limiting null distribution, we can simply use the sum of weighted residuals: $\sum_{j=1}^n\hat \eta_jW(x_j)/n$ with some weight function $W(\cdot)$. This is because under the null, $E(\eta W(X))=0,$ and $\sum_{j=1}^n\hat \eta_jW(x_j)/\sqrt n$ converges  to 
 a normal distribution $N(0,\sigma^2)$ where $\sigma^2$ is easy to obtain.
 This simple  test can fulfill  feature $f_1$) with good performance on the critical value determination and significance level maintenance because it involves fewer unknowns in the limiting variance $\sigma^2$, thus its estimation error would be less.
However,  no researcher would simply use this moment-based test in practice as it is obviously not omnibus and not powerful at all for general alternative models. Further, notice that the conditional moment of $\eta$ given $X$ has the property:  for a nonnegative weight function $\tilde{w}(X)$, $E(E(\eta\cdn X) \tilde{w}(X))^2=0$ and $>0$ under the null  and  alternatives respectively. It can be estimated by, say, the Nadaraya-Watson kernel estimation to define a local smoothing test that is omnibus and  sensitive to oscillating alternative models. This leads to a typical nonparametric estimation-based test, see \cite{zheng1996consistent}. Then feature $f_2$) is achieved. But as we mentioned before, it has slow rate of convergence and suffers from the curse of dimensionality, see \cite{guo2016model} for  more details.  Both have some very obvious shortcomings and  none of these tests can achieve feature $f_3$). But we still stick to these two tests to see how to construct an adaptive-to-model hybrid that shares all appealing features $f_1$) -- $f_3$).  More importantly, the hybrid can make, under the alternatives,  the above nonparametric conditional moment-based component  also share the properties of global smoothing tests satisfying feature $f_3$). This is a somewhat surprising property that will be stated later.

For this mission, we need assistance from other information. Recall that $q = dim(S_{\eta\cdn X})$ is $0$ under the null and is greater than 0 under the alternatives.
Consider a hybrid of weighted moment and conditional weighted moment of the residual $\eta$ in the following format:
\begin{eqnarray}\label{population}
	V = E\{ \eta w(X)[E(\eta w(X)) I(q=0) + E(\eta\cdn X) \tilde{w}(X)I(q>0)] \},
\end{eqnarray}
where $w(\cdot)$ and $\tilde w(\cdot)$ are two weight functions to be determined later.  Thus, under the null, the term $V = E^2[ \eta w(X) ] = E^2(\eta) E^2(w(X)) = 0$ and under the alternatives  $V=E(E^2(\eta\cdn X) \tilde{w}(X){w}(X))= E\{ [E(\Delta(X,\v)\cdn X)]^2 f(X) \}>0$ when we choose $\tilde w=f/w$ where $f$ is the density function of $X$.
Theoretically, the two weight functions can be relatively arbitrarily chosen, but we do have some preferences in practice for this choice, see, e.g. \cite{zheng1996consistent}. When the first component is estimated by the average of weighted residuals and the second component is estimated by a nonparametric method, the constructed test can satisfy features $f_1$) and $f_2$).
Therefore, at the sample level,
an estimator of $V$ in \eqref{population} is defined as
\begin{eqnarray}\label{Vn}
  V_n = \left| \left[ \frac{1}{n}\sum_{j=1}^{n} \hat\eta_j w(x_j) \right]^2 I(\hat q = 0)
        + \frac{1}{n(n-1)} \sum_{j=1}^n \sum_{k\neq j,k = 1}^n\hat\eta_j \hat\eta_k \frac{1}{h^p} K_h(x_j - x_k) I(\hat q>0) \right| ,
\end{eqnarray}
where $K_h(\cdot) = K(\cdot / h)$, $K(\cdot)$ is the kernel function, $h$ is the corresponding bandwidth, and $\hat q$ is the estimated indicative dimension.
Here taking the absolute value is just to keep the non-negativity of $V$ at the sample level.
To get a concise expression, denote
\begin{eqnarray*}
  V_0 = \frac{1}{n}\sum_{j=1}^n \hat \eta_j w(x_j) \quad \text{and }
  V_1 = \frac{1}{n(n-1)}\sum_{j=1}^n\sum_{k\neq j,k=1}^n  \hat \eta_j \hat \eta_k \frac{1}{h^p} K_h(x_j-x_k).
\end{eqnarray*}
Thus $V_n = | V_0^2 I(\hat q = 0) + V_1 I(\hat q >0) | $.

It is clear that to satisfy both features $f_1$) and $f_2$), the estimator $\hat q$ must also be $0$ and $>0$ with a probability going to $1$ respectively under the null and alternatives. We will discuss this property in Section~\ref{sec: dimension estimation} via sufficient dimension reduction. But a very important thing is about the choice of  standardizing constant for $V_n$ in \eqref{Vn} such that the test can also share feature $f_3)$. A standardizing constant $b_n$ must be used such that $T_n=b_nV_n$ has a tractable limiting null distribution. As under the null $V_n=V_0^2$, $b_n=n/\hat \sigma^2_0$ should be used where $\hat \sigma^2_0$ is an estimator of the limiting variance such that $b_nV_0^2$ can be asymptotically chi-square. Under the alternatives this $b_n$ diverges to infinity  faster than $nh^{p/2}$ \citep{zheng1996consistent} or $nh^{1/2}$ \citep{lavergne2012one,guo2016model}. Therefore, $b_nV_n=b_nV_1$ can be much powerful than  \cite{zheng1996consistent}'s test $nh^{p/2}V_1$ or the tests $nh^{1/2}V_1$ if $V_1$ is the test in \cite{lavergne2012one} or \cite{guo2016model}.   This is very unique merit of this adaptive-to-model hybrid. This discussion gives us the idea about how the constructed test could fulfill feature $f_3$).   We will give more details about  the asymptotic properties of the test in Section~\ref{sec: asymptotics}.

  We now determine which estimator of $\sigma_0^2$ we should use. As  $V_n = V_0^2$ under the null where $V_0$ has simple structure and its limiting variance $\sigma_0^2$ has fewer unknowns, we then use it.
For the simplicity of notation, let $E(w^2) = E(w(X)^2)$, $E(\dot g w) = E(\dot g(X,\theta^\ast) w(X) ) $ and $E(\dot g\dot g^\t	) = E(\dot g(X,\theta^\ast) \dot g(X,\theta^\ast)^\t )$.
Under the null, we have
\begin{eqnarray}
  \sigma_0^2 = \sigma_\v^2 [E(w^2)-E(\dot g w)^\t E(\dot g\dot g^\t)^{-1}E(\dot g w)],
\end{eqnarray}
where $\sigma_\v^2 = Var(\v)$.
Write $w_j = w(x_j)$ and $\dot g_{j,n} = \dot g(x_j,\hat \theta_n)$.
A consistent estimator of $\sigma_0^2$ is
\begin{equation}
  \hat\sigma_0^2 = \left( \frac{1}{n}\sum_{j=1}^n \hat \eta_j^2 \right) \left[ \frac{1}{n}\sum_{j=1}^n w_j^2-(\frac{1}{n}\sum_{j=1}^n \dot g_{j,n} w_j)^\t (\frac{1}{n}\sum_{j=1}^n\dot g_{j,n}\dot g_{j,n}^\t)^{-1}(\frac{1}{n}\sum_{j=1}^n \dot g_{j,n} w_j) \right].
\end{equation}
Then, the resulting test statistic is  defined as
\begin{eqnarray} \label{test statistic}
  T_n = \frac{n V_n}{\hat \sigma_0^2}=\frac{n V_0^2I(\hat q=0)}{\hat \sigma_0^2}+\frac{n V_1I(\hat q>0)}{\hat \sigma_0^2}.
\end{eqnarray}
The weight function is chosen as $w(x) = c \exp(-\| x \|)$ for some constant $c>0$ to be specified in the simulations, where $\| \cdot \|$ stands for the Euclidean norm.

\begin{rem}
  We comment on the construction in two aspects.\\ 1). Any nonzero function is applicable as the weight function theoretically, but empirically, we find that a ``small" weight helps enhance the power performance in finite sample scenarios. In theory, a natural question is whether there is an optimal choice of weight function. However, the optimality here is hard to define. For instance, if the optimality is on  the magnitude of limiting variance of $V_0$, the standardization removes the scale.  If the optimality is about the power performance of the test, it relates to the issue about how to construct a most powerful test in a certain sense, which is beyond the scope of this paper.  Thus, we do not discuss this issue in more detail in this paper. \\ 2). Another issue is about choosing an estimator for the limiting variance of $T_n$ as $T_n$ involves two tests in effect. We now explain the use of $\hat \sigma_0^2$.  Let $\Sigma_{Zh}$ be the limiting variance of $ nh^{p/2}V_1$. From  Lemma 3.3 of \cite{zheng1996consistent}, we can see that $\sigma_0^2$ can be smaller than $ \Sigma_{Zh}$ even under the null. In other words, if we use an estimator of $ \Sigma_{Zh}$ in lieu of $\sigma_0^2$, the limiting distribution under the null cannot be a standard chi-square and the values of $T_n$ under the alternatives can be smaller to lower the power.
\end{rem}

\section{Estimation of the indicative dimension} \label{sec: dimension estimation}

In this section, we propose a method  to estimate the dimension $q$. To this end,
 we  define a target matrix with $q$ nonzero eigenvalues below.

\subsection{Target matrix}

 As $\eta$ needs to be replaced by its estimator, the estimation of target matrix and the  dimension $q$ of the corresponding central subspace becomes more complicated. We then consider the following target matrix to make  the  estimation  relatively easier although existing methods such as sliced inverse regression \citep{li1991sliced} and sliced average variance estimation \citep{CookWeisberg1991comment} could also be used.
Denote $A^H$ as the conjugate transpose of a matrix $A$ and define the target matrix as
\begin{eqnarray} \label{target matrix}
  M = \int E[X \exp(it\eta)] E^H[X \exp(it\eta)] dF_\eta(t),
\end{eqnarray}
where $i=\sqrt {-1}$ is the complex number.
For  ease of illustration, assume $E(X) = 0$ and write $Var(X) = \Sigma$.
Let $B$ be a basis of $S_{\eta\cdn X}$ and $P_B(\Sigma) =  B(B^\t \Sigma B)^{-1} B^\t \Sigma$.
The linearity condition is $E(X\cdn B^\t X) = P_B^\t(\Sigma) X$, see condition 3.1 in \cite{li1991sliced}, which is widely used in sufficient dimension reduction.
It holds for $X$ that follows  symmetric elliptical distribution.
\begin{lem} \label{lem: M in central subspace}
 Under the linearity condition,  $Span\{M\} \subseteq \Sigma S_{\eta\cdn X} $. If the rank of $M$ is $q$, then $Span(M) = \Sigma S_{\eta\cdn X}$.
\end{lem}

Further, we have $q = rank(M) = dim(S_{\eta\cdn X})$ since $dim(\Sigma S_{\eta\cdn X}) = dim(S_{\eta\cdn X})$.
Let $\lambda_1\geq \dots\geq \lambda_p\geq 0$ be the ordered eigenvalues of $M$ defined in (\ref{target matrix}).
Then under the null hypothesis, the independence between $\eta$ and $X$ leads to  $\lambda_1 = \dots = \lambda_p = 0$ and $q=0$.
In contrast, under the alternatives, $\lambda_1 \geq \dots\geq \lambda_q > 0 = \lambda_{q+1} = \dots = \lambda_p$ for some $1\leq q \leq p$.

\subsection{Estimation}\label{subsec: estimation of target matrix}

Consider  estimating the target matrix first.
Define $m(t) = E(X\exp(it\eta)) $.
An estimator of $m(t)$ is
\begin{eqnarray}
  \hat m_n(t) = \frac{1}{n} \sum_{j=1}^n x_j \exp(it\hat\eta_j),
\end{eqnarray}
where $\hat\eta_j = y_j - g(x_j, \hat\theta_n)$. The
corresponding estimator of $M$ is
\begin{eqnarray}
  \hat M_n = \frac{1}{n} \sum_{j=1}^n \hat m_n(\hat\eta_j) \hat m_n^H (\hat\eta_j).
\end{eqnarray}
To study this, we first give the consistency of $\hat\theta_n$ in the following lemma.
\begin{lem}\label{lem: consistency of theta_n}
Under the regularity conditions in  Appendix, we have $\| \hat \theta_n - \theta^\ast \| = O_p(n^{-1/2}) $ and
  \begin{eqnarray}
    \hat \theta_n - \theta^\ast = G^{-1} \frac{1}{n} \sum_{j=1}^n \eta_j \dot g(x_j,\theta^\ast) + o_p(\frac{1}{\sqrt n}),
  \end{eqnarray}
  where $G = E[\dot g(X,\theta^\ast) \dot g^\t (X,\theta^\ast) ] $.
\end{lem}
This asymptotically linear representation is a standard result for the least squares estimation, see \cite{zheng1996consistent} for details.

Let $X_i$ be the $i$-th component of $X$ and $\|\cdot \|$ be the Frobenius norm. The following theorem states the consistency of $ \hat M_n $.

\begin{thm} \label{theorem: consistency of Mn}
  Suppose $\max_{1\leq j\leq p} E(|X_j|^4)<\infty $. Then
  \begin{enumerate}
    \item under the null hypothesis, $ \| \hat M_n - M \|  = O_p(n^{-1}) $;
    \item under the alternative hypothesis, $ \| \hat M_n - M \|  = O_p(n^{-1/2}) $.
  \end{enumerate}
\end{thm}

These results provide a base for estimating the indicative dimension $q$.
We then suggest a criterion that is a slight modification of the thresholding double ridge ratio (TDRR, hereafter) method developed by \cite{zhu2016dimensionality}.
Define the eigenvalues of $\hat M_n$ to be $\hat\lambda_1\geq \dots \geq \hat\lambda_p\ge 0$ and $\hat s_j = \hat\lambda_j / (\hat \lambda_j +1)$.
We then use the following two-step ratio to determine the estimation of dimension $q$. Let
\begin{eqnarray}
\hat s_j^\ast = \frac{\hat s_j^2 + c_{1n}}{\hat s_{j+1}^2 + c_{1n}} -1 \quad \textit{ and } \hat r_j =  \frac{\hat s_{j+1}^\ast + c_{2n}}{\hat s_j^\ast+c_{2n} },
\end{eqnarray}
where $c_{1n}$ and $c_{2n}$ are the two ridges that converge to $0$ in proper rates to be selected later, such that the dimension $q$ can be identified.
The criterion for the determination of $q$ is
\begin{eqnarray}
\hat q = \begin{cases}
0, & \text{ if $\hat r_j > \tau$ for all $j \in \{ 1, \dots, p\}$ }, \\
\arg\max\limits_{1\leq j\leq p} \left\{ j:  \hat r_j \leq \tau \right\}
\end{cases},
\end{eqnarray}
with a threshold $0< \tau <1 $.
Based on the rule of thumb as in \cite{zhu2016dimensionality}, we also set $\tau = 0.5$.
Details of the modified TDRR method and corresponding proofs are shown in the supplementary materials.


Following theorem states the consistency of $\hat q$ to indicate the underlying model.

\begin{thm} \label{theorem: consistency of q}
Under the conditions in Appendix, we have
  \begin{enumerate}
    \item if $c_{1n}\to 0 $, $c_{2n}\to 0 $ and $n^2 c_{1n}c_{2n} \to \infty $, then under the null hypothesis of (\ref{Hypo_0}), $\pr( \hat q = 0 ) \to 1 $;
    \item if $c_{1n}\to 0 $, $c_{2n}\to 0 $ and $n c_{1n}c_{2n} \to \infty $, then under the alternative hypothesis of (\ref{Hypo_1}), $\pr( \hat q =q>0 ) \to 1 $;
  \end{enumerate}
\end{thm}

{ 
Due to the different properties of the target matrix from those in \cite{zhu2016dimensionality}, the ridges are chosen differently from the ones used in their method.
Based on our experience in the numerical studies, the recommended ridges are $c_{1n} = 3e^{-4}\sqrt 8 \log n / \sqrt n $ and $c_{2n} = 4\sqrt 8 \log n/5\sqrt n$.
}

\section{Asymptotic properties} \label{sec: asymptotics}

%

\subsection{Asymptotics under different hypotheses}

Theorem~\ref{theorem: consistency of q} offers a very important result such that the proposed test is of the following asymptotic model adaptation property.

\begin{lem} \label{lemma: asymptotic equivalence}
 Assume the conditions in Appendix. Then as $n\to \infty$, with a probability going to $1$, under the null hypothesis of \eqref{Hypo_0}, $ T_n =nV_0^2/\hat \sigma^2_0$ and under the alternative hypothesis of \eqref{Hypo_1}, $ T_n =n|V_1|/\hat \sigma^2_0$.
\end{lem}

From this lemma, 
we first give the result about the limiting null distribution.

\begin{thm} \label{thm: limiting null distribution}
  Under the null hypothesis of \eqref{Hypo_0} with the regularity conditions in  Appendix, the test statistic satisfies
  \begin{equation*}
    T_n \to \chi_1^2
  \end{equation*}
  in distribution where $\chi_1^2$ stands for chi-square distribution with one degree of freedom.
\end{thm}

Now we investigate the power performance of the test under the alternative hypothesis of \eqref{Hypo_1}.
Recall that $\eta = \Delta(X,\v) $.
For notational simplicity, rewrite $\Delta(X,\v)$ as $\Delta$,  $f(X)$ as $f$, $\dot g(X,\theta^\ast)$ as $\dot g$ and  $w(X)$ as $w$ hereafter.
Denote $\Sigma_{01}= E(\Delta^2)[E(w^2)-E(\dot g w)^\t E(\dot g \dot g^\t)^{-1}E(\dot g w)]$ and $\mu = E(\Delta^2 f)/ \Sigma_{01}$.
We have the following result.

\begin{thm} \label{thm: limiting under alternatives}
  Given the regularity conditions in  Appendix, under the alternative hypothesis of \eqref{Hypo_1}, in probability,
  \begin{equation*}
    T_n/n \to \mu >0.
  \end{equation*}
\end{thm}

%
Now we consider the sequence of local alternative models:
\begin{equation} \label{H_1n}
    Y = g(X,\theta_0)+\delta_n \ell(X)+\v,
\end{equation}
where $\delta_n\to 0$ as $n\to \infty$ and $\ell(\cdot)$ is a non-constant function.

Following lemma states the consistency of $\hat\theta_n$ under such local alternatives.

\begin{lem} \label{lem: consistency of theta under local}
  Under the local alternatives in (\ref{H_1n}) and the conditions in Appendix, we have
  \begin{eqnarray}
    \hat\theta_n - \theta_0 = G^{-1} \frac{1}{n} \sum_{i=1}^n \v_i \dot g_i + \delta_n G^{-1} E(\dot g \ell) + o_p(\frac{1}{\sqrt n}).
  \end{eqnarray}
\end{lem}

Before presenting the asymptotic result of the test, we give some results about the estimator $\hat q$ under the local alternatives  to show why we call the dimension the indicative dimension.

{
\begin{thm} \label{thm: q under local}
Assume that the conditions in  Appendix hold, then under the local alternative model of \eqref{H_1n},
\begin{enumerate}
    \item if $\delta_n = n^{-1/2}$, $c_{1n}\to 0 $, $c_{2n}\to 0 $ and $n^2 c_{1n}c_{2n} \to \infty $,
    then $\pr (\hat q = 0)\to 1$;
    \item if $\delta_n = n^{-\alpha}$ for some $0<\alpha<1/2$, $c_{1n}= o_p(\delta_n^4) $, $c_{2n}\to 0 $, $ n c_{1n}c_{2n} / \delta_n^2 \to \infty $ and $c_{1n}c_{2n}/\delta_n^6 \to\infty $,
    then $\pr (\hat q =1)\to 1$.
\end{enumerate}
\end{thm}

\begin{rem}
  Theorem \ref{thm: q under local} provides two pieces of important information on the model indication through $\hat q$. First,  for any rate of $\delta_n$ as close to $1/\sqrt n$ as possible, $\hat q=1$ cannot be equal to the true dimension $q$, but can still well indicate the local alternatives. This rate is the fastest possible rate for such a separation in this research area as when $\delta_n$ goes to zero at or faster than $1/\sqrt n$, $\hat q$ loses the indication ability to the local alternatives.  In other words,  TDRR can reach the optimal rate $\delta_n$ of convergence while the only existing method proposed by \cite{guo2016model} can only identify the dimension  at the rate such that $n^{-1/2}h^{-1/4}/\delta_n=o(1)$ even under the dimension reduction framework.
 We also note that these results, similarly as those in \cite{guo2016model},  basically have theoretical meaning, unless we have prior information on the closeness of local alternatives to the null, we can then set those values. It deserves further study to make it useful in practice.
 \end{rem}
}

Based on these results, we are now in the position to discuss the asymptotic properties of the proposed test in detail.
Define $\mu_0 = E(\ell w)-E(\dot g w)^\t E(\dot g \dot g^\t)^{-1} E(\dot g \ell) $, $\mu_1 = E(\ell^2 f)/\sigma_0^2$, $ \Sigma_1 = 2 \sigma_\v^2 \int K^2(u)du \int f^2(x)dx $ and $Var = \Sigma_1/\sigma_0^2$. We state the power performance of the test in various scenarios.

\begin{thm} \label{thm: asymptotic under local alternatives}
  Given conditions in  Appendix with the ridges $c_{1n}$ and $c_{2n}$ satisfying the conditions in Theorem~\ref{thm: q under local}, if $h \to 0$ and $nh^p \to \infty$, then under the local alternatives of \eqref{H_1n},
    \begin{enumerate}
      \item if $\delta_n = n^{-1/2}$,
      $T_n \to \chi_1^2( \mu_0^2/\Sigma_0)$
      in distribution;
      \item if $\delta_n = n^{-\alpha}$, $ 0< \alpha < 1/2 $, and
        \begin{enumerate}
          \item $ n^{1/2}h^{p/4}\delta_n\to 0 $,  $T_n/h^{-p/2} \to
          |N(0,Var)| $ in distribution,
          \item $\delta_n = n^{-1/2}h^{-p/4} $,   $T_n/h^{-p/2} \to |N(\mu_1,Var)| $ in distribution,
          \item $ n^{1/2}h^{p/4}\delta_n \to\infty $,  $T_n/(n\delta_n^2) \to \mu_1 $ in probability.
        \end{enumerate}
    \end{enumerate}
\end{thm}

\begin{rem}
 Together with Theorem~\ref{thm: limiting under alternatives}, the results are  informative and important which shows that feature $f_3$) can be achieved. As commented in Remark~1,  \cite{zheng1996consistent}'s test and the tests in \cite{lavergne2012one} and \cite{guo2016model} diverge to infinity at the nonparametric rate of order $nh^{p/2}$ and of $nh^{1/2}$.  Theorem~\ref{thm: limiting under alternatives} shows that although under the global alternative $T_n$ uses the lobal smoothing test proposed by \cite{zheng1996consistent} up to the standardizing constant, it can
 diverge to infinity at the fastest possible rate of order $n$.
Further, result 1) shows that the nonparametric estimation-based  component $V_1$ is used, $T_n$ can still detect the local alternatives distinct from the null at the fastest possible rate $1/\sqrt n$. Result 2) gives a full picture to show that when $\delta_n$ converges to zero  slower, the test statistic can have different weak limits at different rates. We can see that in case (b), $T_n$ diverges to infinity at the rate of $h^{-p/2}$ whereas \cite{zheng1996consistent}'s test goes to {\color{red} a finite limit}.
 All these results demonstrate that the special construction of the hybrid makes the nonparametric estimation-based component behave like a global smoothing test. This is a significant contribution of the method.
 The numerical studies reported later will  also justify the  power performance improvement over existing local smoothing tests.
 \end{rem}
%

\section{Numerical studies}\label{sec:NumericalStudy}

\subsection{Simulations}
In this section, we conduct several simulation studies to examine the  performance of the proposed test through  comparisons with several typical local and global smoothing tests in different scenarios.

The proposed  test is written as $T_n$. The typical local smoothing tests are compared: the  kernel estimation-based test $T^{Zh}$ proposed by  \cite{zheng1996consistent} and the kernel estimation-based adaptive-to-model test $T^{GWZ}$  in \cite{guo2016model}.
Here we make a slight modification of the \cite{guo2016model}'s test such that it can be applied to the general parametric model considered in this paper other than the parametric single-index model in \cite{guo2016model}.
To be precise, assume $q_1 = dim(S_{Y\cdn X})$ and $Span\{ B_1 \} = S_{Y\cdn X} $ with the rank of $B_1$ equal to $q_1$.
Without confusion, we still write the modified \cite{guo2016model}'s test as $T^{GWZ}$ in the following:
\begin{eqnarray}
  T^{GWZ} = n h^{\hat q_1/2} \frac{1}{n(n-1)} \sum_{j=1}^n \sum_{k\neq j, k=1}^n \hat\eta_j \hat \eta_k \frac{1}{h^{\hat q_1}} K_h(\hat B_1^\t (x_j-x_k)) ,
\end{eqnarray}
where $\hat q_1$ and $\hat B_1$ are the corresponding estimate of $q_1$ and $B_1$.
In the experiments, we use the DEE-SIR method and the BIC-type criteria used in \cite{guo2016model} to estimate the target matrix $B_1$ and its rank $q_1$.
Furthermore, we also compare it with two typical global smoothing tests: the ICM test $T^B$ proposed by  \cite{bierens1982consistent} and the empirical process-based test $T^{SMQ}$ proposed by \cite{stute1998bootstrap}.

We want to provide the following information on the performance. First, note that in our test, the second component $V_1$ is exactly the same as that in \cite{zheng1996consistent} except for different standardizing constant $n/\hat \sigma_0^2$.  \cite{guo2016model} has well demonstrated that \cite{zheng1996consistent} suffers from the curse of dimensionality severely and $T^{GWZ}$ fully utilizes the dimension reduction structure with much better power performance  than $T^{Zh}$ when the dimension is high. Thus, through the comparison with $T^{GWZ}$, we want to check how the hybrid can well overcome the shortcomings $T^{Zh}$ has. Second, we choose two global smoothing tests to compare as we want to see how the hybrid can share the nice features of global smoothing tests  under low-frequency alternatives and can acquire higher power under high-frequency alternatives. Based on these considerations, we design the following studies.

Study 1 and Study 2 consider some nonlinear null models  against high-frequency and low-frequency alternative model respectively.
In Study 1, we also examine the influence of dependency among the components of $X$ to the performances of the competitors.
These  studies  focus on the scenarios with $q=1$ under the alternatives.
In Study 3, we consider a model with $q=p-1$ under the alternatives.
The null models in these $3$ studies are all parametric single-index.
To further check the performance of the hybrid test, in Study 4, we design a  model without dimension reduction structure under the null hypothesis.
Every experiment in the simulations is repeated  $1,000$ times to compute the empirical sizes and powers.
The significance level is set to be $0.05$.

\begin{figure}[htbp]%
    \centering
    \subfloat[$p=2$]{{\includegraphics[width=.65\linewidth]{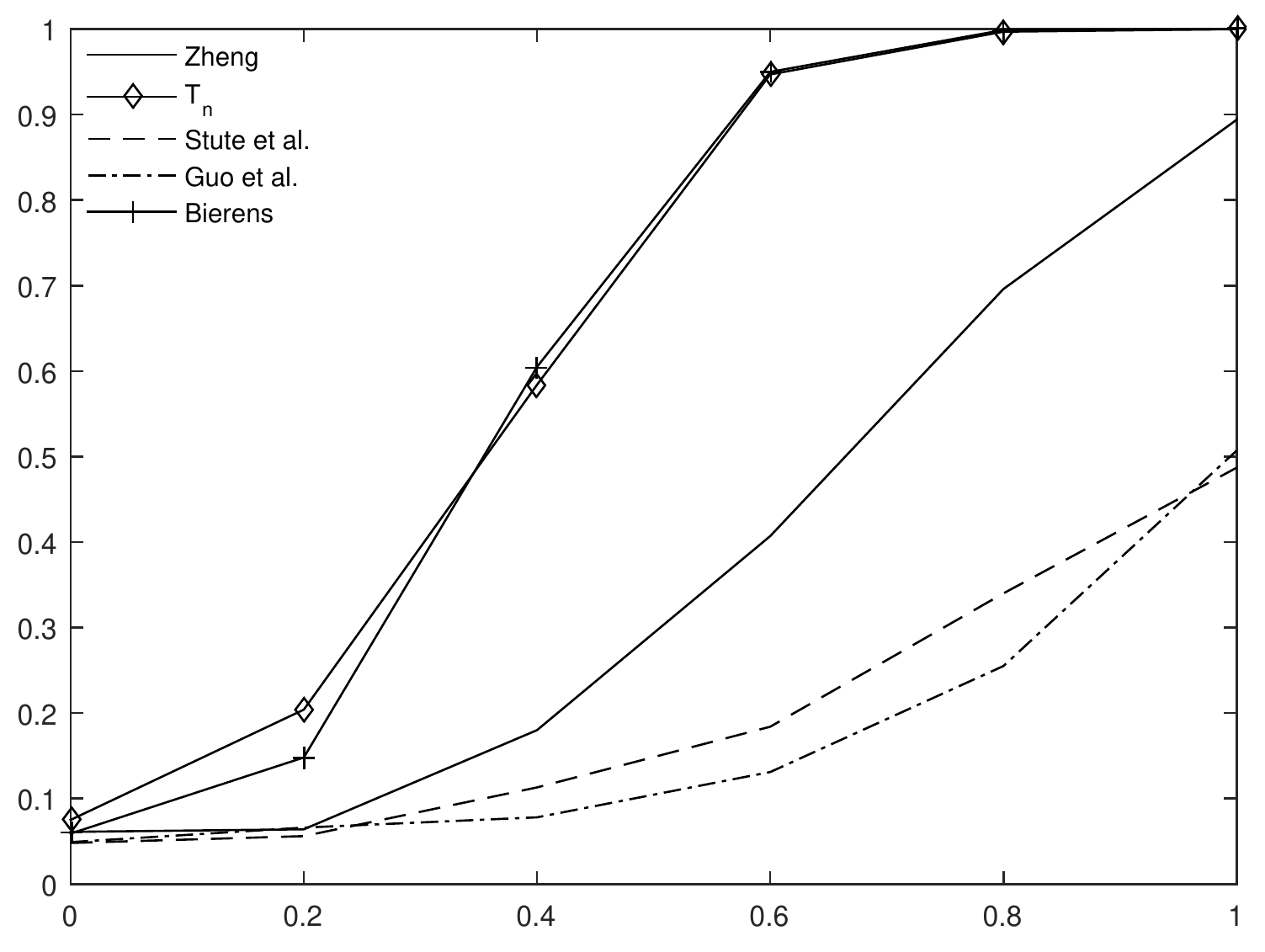} }}%
    \qquad
    \subfloat[$p=8$]{{\includegraphics[width=.65\linewidth]{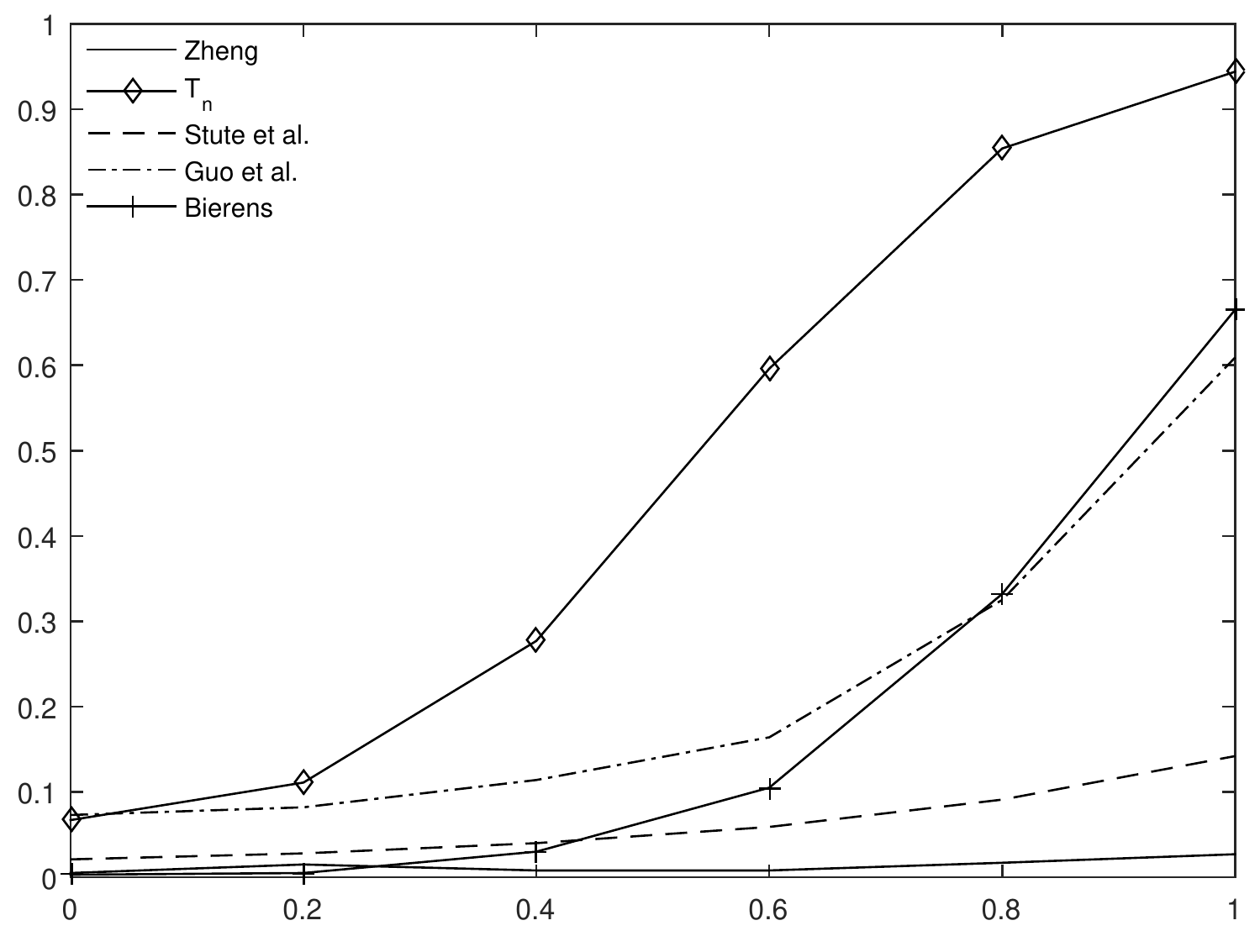} }}%
    \caption{Empirical sizes and powers with $X\sim N(0,I_p)$ in Study $1$} \label{fig:exp_sin_independent X}
\end{figure}

{\bf Study 1:}\,
Consider a   model as:
\begin{eqnarray*}
  Y = 0.25 \exp(2 \alpha^\t X) + a \sin(\beta^\t X) + \v.
\end{eqnarray*}
The parameters are $\alpha = (1,\dots,1, 0,\dots, 0)/ \sqrt{p/2} $ and $\beta = (0,\dots, 0, 1,\dots,1)/ \sqrt{p/2} $, where $\|\alpha\| = \|\beta\| = 1$.
The value $a=0$ corresponds to the null hypothesis and $a\not =0$ to the alternatives. The sample size $n = 200$ and $p=2, 8$ to respond low and high-dimensional scenarios respectively. We do not consider higher dimension only because when $p=8$, local smoothing tests have already clearly shown their poor power performance as \cite{guo2016model} demonstrated.
We consider two different distributions of $X$ with independent and dependent components: $N(0,I_p)$ and $N(0,\tilde \Sigma)$ where the $(i,j)$-th element of $\tilde \Sigma$ is $0.5^{|i-j|}$.
Empirical sizes and powers with independent $X$ are shown in Figure~\ref{fig:exp_sin_independent X}.
Table~\ref{tab:exp_sin_dependent X} displays the results with $X\sim N(0,\tilde \Sigma)$.

\begin{table*}[htbp]
  \centering
  \caption{Empirical sizes and powers with $X\sim N(0,\tilde \Sigma)$ in Study $1$}
    \begin{tabular}{c cccccc}
    \toprule
    \toprule
    $p=2$   & $a$     & \multicolumn{1}{c}{$T^{Zh}$} & \multicolumn{1}{c}{$T^{n}$} & $T^{SMQ}$ & $T^{GWZ}$ & $T^B$ \\
    \midrule
          & 0     & 0.062 & 0.067 & 0.050 & 0.067 & 0.049 \\
          & 0.2   & 0.077 & 0.205 & 0.136 & 0.088 & 0.172 \\
          & 0.4   & 0.197 & 0.631 & 0.426 & 0.211 & 0.598 \\
          & 0.6   & 0.438 & 0.944 & 0.755 & 0.486 & 0.956 \\
          & 0.8   & 0.743 & 0.999 & 0.938 & 0.833 & 1.000 \\
          & 1     & 0.934 & 1.000 & 0.999 & 0.985 & 1.000 \\
    \midrule
    \midrule
    $p=8$   & $a$     & \multicolumn{1}{c}{$T^{Zh}$} & \multicolumn{1}{c}{$T^{n}$} & $T^{SMQ}$ & $T^{GWZ}$ & $T^B$ \\
    \midrule
          & 0     & 0.018 & 0.070 & 0.041 & 0.059 & 0.009 \\
          & 0.2   & 0.024 & 0.105 & 0.110 & 0.086 & 0.025 \\
          & 0.4   & 0.025 & 0.268 & 0.291 & 0.101 & 0.198 \\
          & 0.6   & 0.047 & 0.534 & 0.595 & 0.159 & 0.649 \\
          & 0.8   & 0.053 & 0.779 & 0.801 & 0.243 & 0.943 \\
          & 1     & 0.075 & 0.896 & 0.915 & 0.384 & 0.997 \\
    \bottomrule
    \end{tabular}%
  \label{tab:exp_sin_dependent X}%
\end{table*}%


For $X$ with independent components, when $p=2$, the hybrid test and \cite{bierens1982consistent}'s test perform best and when $p=8$, our test shows a great advantage over the other competitors.
For the tests in \cite{zheng1996consistent}, \cite{bierens1982consistent} and \cite{stute1998bootstrap}, their powers are severely deteriorated  by the dimension $p$.
\cite{guo2016model}'s test keeps a stable performance when $p$ goes  from $2$ up to $8$. This makes sense as the model structure is single-index, which is in favor of their test.

For $X \sim N(0,\tilde \Sigma)$, again the proposed test and \cite{bierens1982consistent}'s test are the winners when $p=2$.
Our test is compatible with the global smoothing test in \cite{stute1998bootstrap}, which works best with $p=8$.
Although \cite{bierens1982consistent}'s test has a higher power than \cite{stute1998bootstrap}'s, its empirical size  is far away from the significance level of $0.05$. Thus, the significance level maintenance is an issue for this test.

The results with $X\sim N(0,I_p)$ and $X\sim N(0,\tilde \Sigma)$ suggest that our test is robust to the dependency between components of $X$.
\cite{guo2016model}'s test has a better performance with $X\sim N(0,\tilde \Sigma)$ than $X\sim N(0,I_p)$ when $p=2$.
However, its power drops  quickly with increasing dimension when the components of $X$ are correlated.
\cite{stute1998bootstrap}'test performs much better under $X\sim N(0,\tilde \Sigma)$ than under $X\sim N(0,I_p)$.

The comparison with \cite{zheng1996consistent}'s test clearly shows the advantage of the hybrid even when \cite{zheng1996consistent}'s test is, up to the standardizing constant, its nonparametric conditional comment component.

\begin{figure}[htbp]%
    \centering
    \subfloat[$p=2$]{{\includegraphics[width=.65\linewidth]{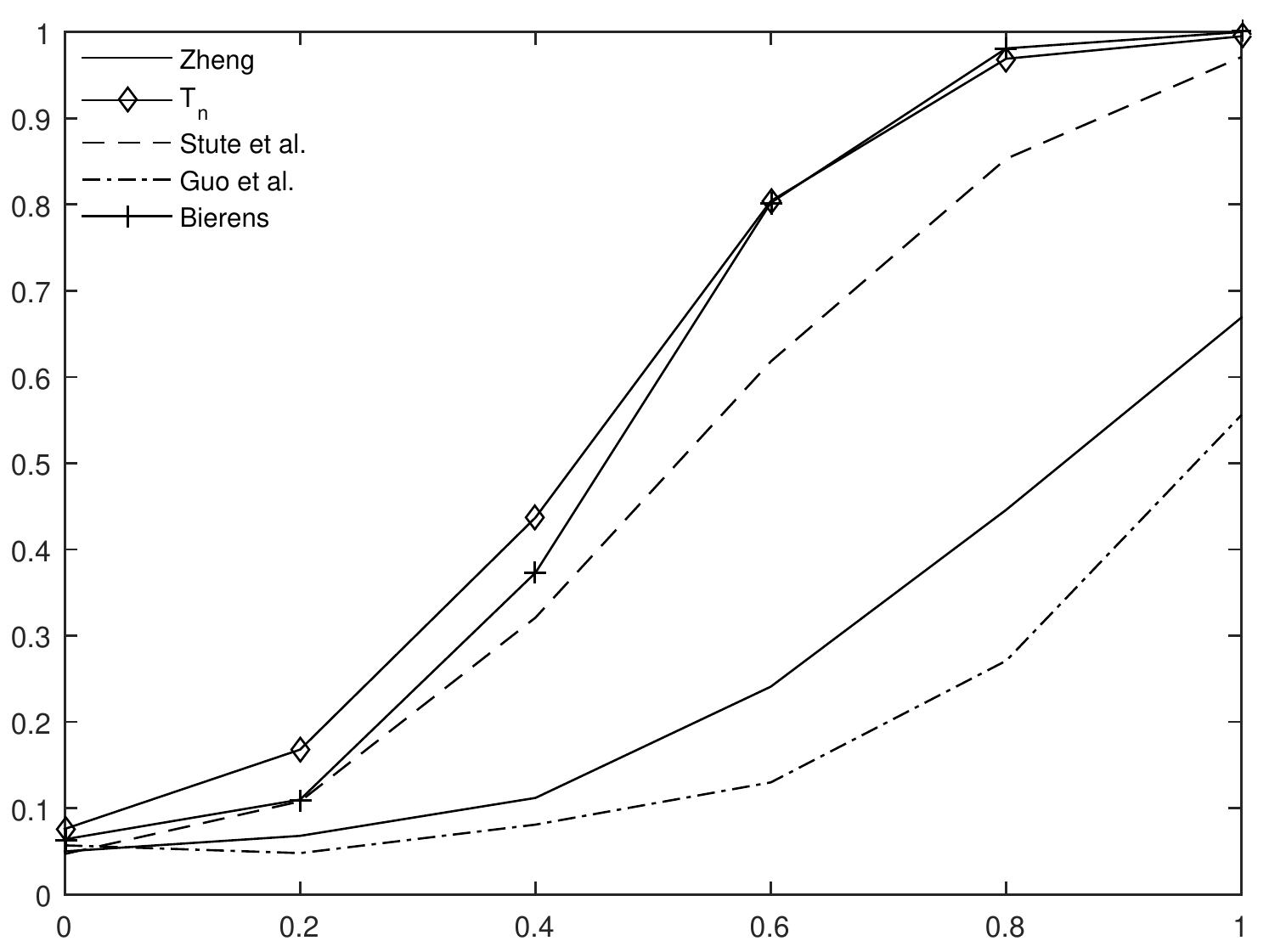} }}%
    \qquad
    \subfloat[$p=8$]{{\includegraphics[width=.65\linewidth]{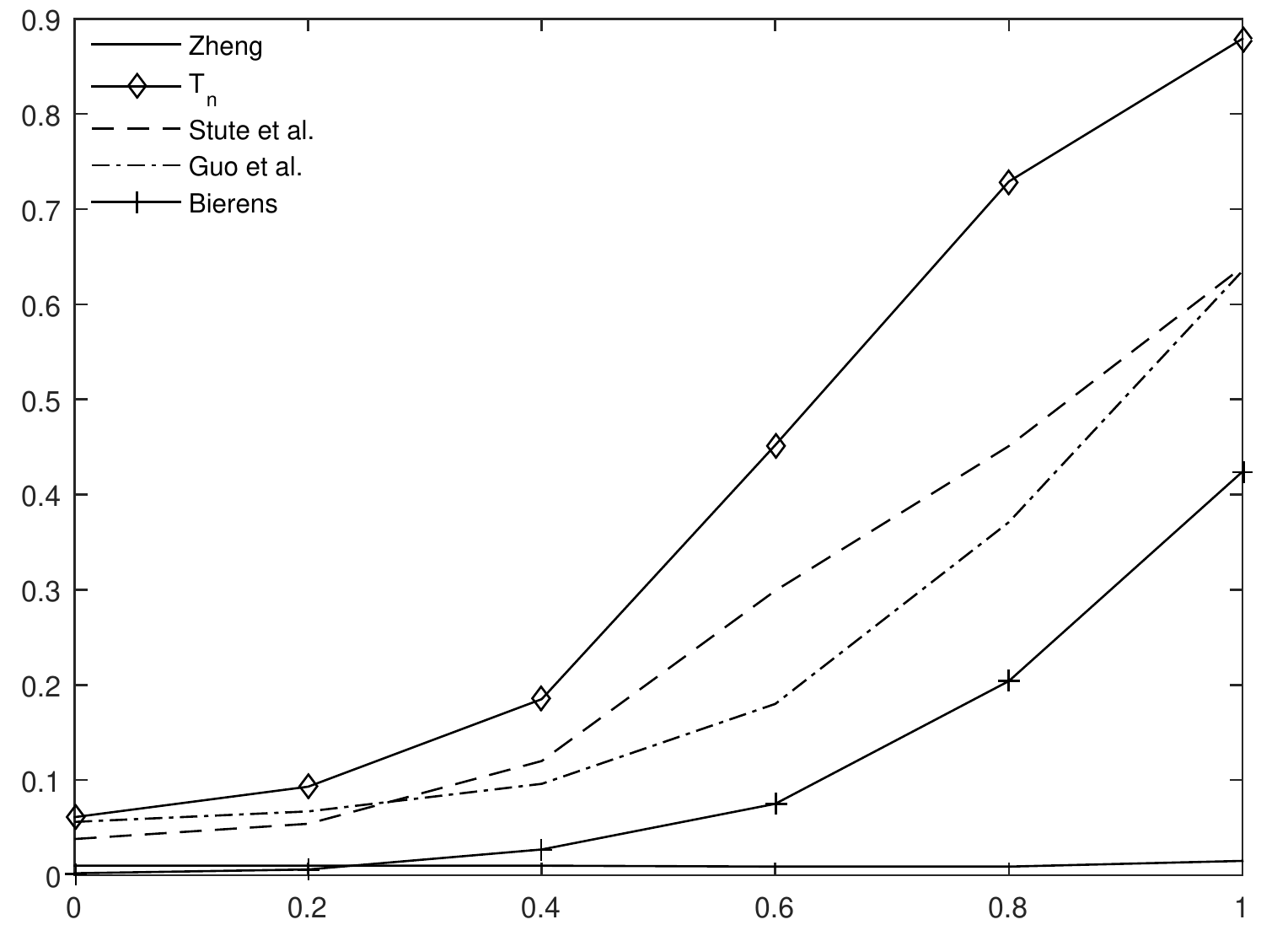} }}%
    \caption{Empirical sizes and powers in Study $2$} \label{fig:LinSqu_tanh}
\end{figure}

{\bf Study 2:}
Consider a nonlinear null model against low-frequency  alternative models:
\begin{eqnarray*}
   Y = \alpha^\t X + 0.8(\alpha^\t X)^2 + a \tanh(\beta^\t X) + \v.
\end{eqnarray*}
The parameters $\alpha$ and $\beta$ follows the previous settings.
 $X$ and  $\v$ are generated from $N(0,I_p)$ and $N(0,1)$ respectively.
Simulation results are presented in Figure~\ref{fig:LinSqu_tanh}.

The hybrid test performs best among the competitors in both cases with $p=2$ and $p=8$.
Except for the proposed test and \cite{guo2016model}'s test, the others suffer from severe power lose due to the dimension increasing.
\cite{bierens1982consistent}'s test is compatible with our test when $p=2$, but deviates far from the significance level  and low power when $p=8$.
\cite{zheng1996consistent}'s test, as discussed before, slowly converges to its weak limit when $p=8$, which may be the cause for low power.
\cite{guo2016model}'s test performs much better than \cite{zheng1996consistent}'s when $p=8$.
Power of \cite{stute1998bootstrap}'s test declines sharply as $p$ increases up to $8$.

{\bf Study 3:}\,
In the above two studies, the indicative dimension $q = 1$ under the alternatives. In this study, we conduct experiments for models with $q=p-1>1$ under the alternatives.
Let $X_i$ be the $i$-th component of $X$.
Consider
\begin{eqnarray*}
   Y = 0.25 \exp(2 X_1) + a \{ 0.5 X_2^3 + \cos(X_3) + X_4 - |X_5| + \tanh(0.6\pi X_6) + X_7 X_8 \} + \v.
\end{eqnarray*}
The sample size $n = 200$ and dimension $p = 8$.
The explanatory vector $X$ and the error $\v$ are drawn independently from $N(0,I_p)$ and $N(0,1)$.
Table \ref{tab:exp_full} presents the empirical sizes and powers.

\begin{table*}[htbp]
	\centering
  \caption{Empirical sizes and powers in Study $3$}
\begin{tabular}{c ccccc}
	\toprule
	\toprule
	$a$     & $T^{Zh}$ & $T_n$ & $T^{SMQ}$ & $T^{GWZ}$ & $T^B$ \\
	\midrule
	0     & 0.013 & 0.066 & 0.016 & 0.069 & 0.002 \\
	0.2   & 0.006 & 0.155 & 0.035 & 0.072 & 0.005 \\
	0.4   & 0.020 & 0.445 & 0.099 & 0.180 & 0.043 \\
	0.6   & 0.020 & 0.661 & 0.109 & 0.439 & 0.130 \\
	0.8   & 0.021 & 0.780 & 0.194 & 0.734 & 0.254 \\
	1     & 0.016 & 0.798 & 0.240 & 0.895 & 0.403 \\
	\bottomrule
    \end{tabular}%
  \label{tab:exp_full}%
\end{table*}%


The results show that the proposed test and \cite{guo2016model}'s test work better than the others.
When the deviation from the null hypothesis is small, which corresponds to $a = 0.2, 0.4, 0.6, 0.8$, the proposed test constantly surpasses the other competitors.
This is consistent with the property that our test can better detect the local alternatives.
Compared with the others, \cite{guo2016model}'s test also works well, and  when $a=1$, its power  even suddenly jumps up to be higher than ours.

Under the null hypothesis, the models in Study 1 - Study 3 are all single-index where $dim(S_{Y\cdn X}) = 1$ under the null hypothesis.
To make the comparison more comprehensive, we consider a null model with $dim(S_{Y\cdn X})>1$ in the following  study.

\

\begin{figure*}[htbp]
  \centering
  \includegraphics[width=0.5\linewidth]{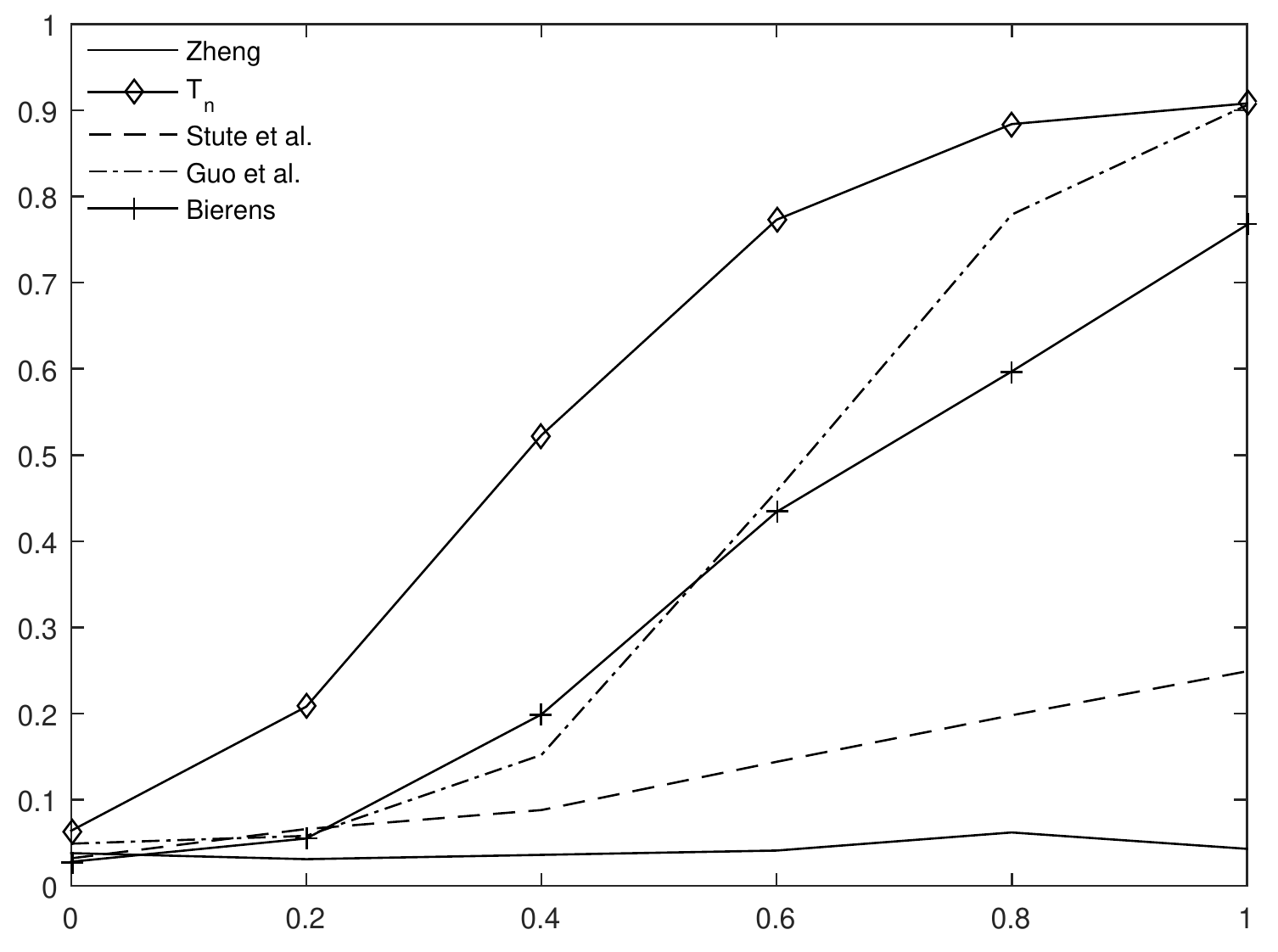}
  \caption{Empirical sizes and powers in Study $4$}
  \label{fig:full_null}
\end{figure*}

{\bf Study 4:}\,
Recall that $X_i$ represents the $i$-th component of $X$.
Consider a null model without a dimension reduction structure:
\begin{eqnarray*}
  Y = c_1 X_1 + c_2 X_2^2 + c_3 X_3^3 + c_4 X_4 X_5 + c_5 \sin(X_6) + a\{ 0.2 X_1^2 + 0.3 X_2^3 \} + \v.
\end{eqnarray*}
The unknown parameters in the null model are $(c_1, c_2, c_3, c_4, c_5) = C/\|C\|$ where $C = (1,1/2,1/3,1,1)$.
Figure~\ref{fig:full_null} displays  the empirical sizes and powers of the all competitors.

From the results, we can see that the proposed test has a great advantage over the competitors, especially under the scenarios with small $a = 0.2, 0.4, 0.6$.
\cite{zheng1996consistent}'s test and \cite{stute1998bootstrap}'s test does not  work well with this model.
The power of \cite{guo2016model}'s test is also close to $1$ when $a = 1$, but its ability to detect the alternatives with small $a$ is not comparable with the proposed test.
Again,
the empirical size of \cite{bierens1982consistent}'s test is too conservative to maintain the significance level.

%
%
%


\subsection{A real data example}
{
The data of real estate price were collected from June 2012 to May 2013, see \cite{yeh2018building}. This data set contains $414$ samples during this period. Let $Y$ be the price per unit area of the interested real estate. {\color{blue}}
Analysis in \cite{yeh2018building} shows that the important attributes involve the date of transaction ($X_1$),
the house age ($X_2$),
the distance to the nearest metro station ($X_3$),
the number of convenience stores within walking distance ($X_4$) and the geographic location ( denoted by northern latitude $X_5$ and eastern longitude $X_6$).
The transaction dates are transformed into real numbers. For instance, if a house was traded in September 17, 2012, then its transaction date is presented as $2012.917$.
 Compared with traditional comparative approaches, regression analysis costs lower and is more reasonable since it reduces the bias caused by subjectivity.
The following regression model was applied to fit this real estate price dataset:
\begin{eqnarray*}
  E(Y|X) &=& \alpha_0+\sum_{j=1}^{6}\alpha_j X_j + \alpha_7 X_5^2 +\alpha_8 X_6^2 + \alpha_9 X_5 X_6.
\end{eqnarray*}
They also compared the advantages and disadvantages of various methods to appraise the real estate price and concluded that the above regression model is then preferred by industry for its interpretability and low cost in large scale evaluation. Now we conduct a test to verify the adequacy of this model.
The value of the test statistic is $1.0249$ and the $p$-value is $0.3114$. Therefore the model is plausible in describing the relationship between the target real estate price and the important attributes.
The residual plot in Figure~\ref{resiplot_TaiwanHouse} shows that there is no obvious nonlinear pattern in the residuals, which also implies this conclusion.
\begin{figure*}[h]
  \centering
  \includegraphics[width=0.5\linewidth]{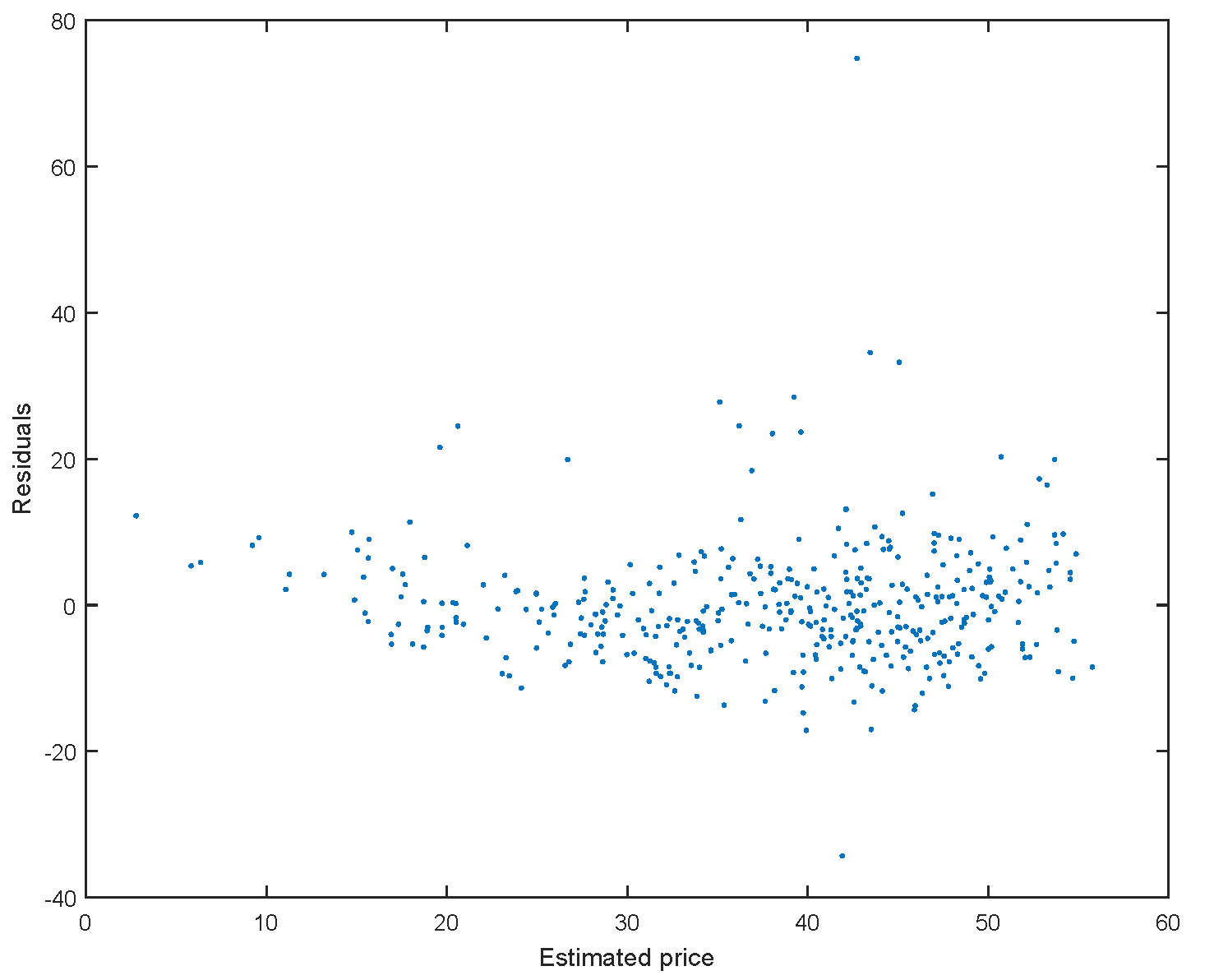}
  \caption{Residual plot}\label{resiplot_TaiwanHouse}
\end{figure*}%

}

\section{Discussions} \label{sec: discussions}

In this research, we propose a hybrid test that can have model adaptation property. This is the first test in this research field, to the best of our knowledge, to inherit all main advantages of nonparametric estimation-based and empirical process-based tests in the literature.  Further, this methodology is ready to be applied to other testing problems with more complicated data.

It is interestingly observed that we use two tests that are definitely not powerful to construct a very powerful test, even more powerful than any existing tests in the sense that it can handle both low-frequency and high-frequency alternative models. Thus, the most significant contribution of this method is that under the alternatives, the hybrid can make a local smoothing test share all appealing features global smoothing tests have. Note that the test under the null is just for critical value determination. Thus, we should use a test that is simple such that the significance level can be easily maintained. For the test under the alternatives, it could be any nonparametric estimation-based test.  A natural question, though beyond the scope of this paper, is whether we can choose a proved powerful test to make the test more powerful. It could be possible because the hybrid can consist of any two tests and thus deserves further study. On the other hand, we also see from the results in Sections \ref{sec: proposed test} and \ref{sec: asymptotics} that for any nonparametric estimation-based test, the asymptotic properties of the resulting hybrid test should be similar, whereas the pair of the tests are moment-based and empirical process-based should make the resulting test only satisfy features~$f_1$ and $f_3$, but not feature~$f_2$ as both are global smoothing tests.

From the proof for Theorem \ref{thm: q under local}, we can find the method of dimensionality determination has the following  property: \,
{\it When the null model is linear, even under the local alternatives (\ref{H_1n}), the estimation $\hat q$ equals $0$ with a probability going to $1$.}\,
In other words, when the null model is linear, the dimension $\hat q$ cannot indicate the local alternatives designed in our paper. We find that this is because of the target matrix construction.
For a linear null model, we have $g(X, \theta)=\theta^\t X$ and thus $\dot g = X$. Together with $m(t) = E(X\exp(it\eta)) $ in Section \ref{subsec: estimation of target matrix}, it will lead to $\Delta_\ell = E(X\ell) - E(X\dot g^\t) E(\dot g \dot g^\t)^{-1} E(\dot g \ell) = 0$, as shown by the proof in the supplementary materials.
To avoid this problem, we may use a more general function $L(X)$ in lieu of $X$ such that $\Delta_\ell = E(L(X)\ell) - E(L(X)\dot g^\t) E(\dot g \dot g^\t)^{-1} E(\dot g \ell) \not = 0$. We point out this problem here, but in this paper we still use the target matrix provided in Section \ref{subsec: estimation of target matrix} for simplicity. How to choose a target matrix such that the dimensionality determination works better deserves future study.

Another issue is about the extension of the method to heteroscedastic models. The current approach has a main limitation that only homoscedastic models can be handled. From the results, we can see that the key to handling heteroscedastic null models is to define an indicative dimension and to have a method to identify it. The method currently relies on the independence between the error and covariates. For heteroscedastic models, the independence  no longer holds. Thus, in a special case where $\varepsilon=b(X)\epsilon$ and $X$ is independent of $\epsilon$, we may estimate $b$ and then use $\eta/\hat b$ instead of $\eta$ to construct the test. For more general paradigms, it deserves further study.

\bigskip
\begin{center}
	{\large\bf SUPPLEMENTARY MATERIAL}
\end{center}

\begin{description}
	
	\item[Supplemenraty of adaptive-to-model hybrid test for regressions] Technical details of TDRR and proofs of the theorems. (.pdf file)

\end{description}

\appendix
\setcounter{secnumdepth}{0}
\section{Appendix}
\label{appendix}

\subsection*{Regularity conditions}
\begin{assm}
  $\{(x_i,y_i)\}_{i=1,\dots,n} $ are i.i.d. random samples from $(X,Y) $ in $R^p\times R $ and $EY^2<\infty $.
\end{assm}

\begin{assm}
  The  parameter space $\Theta$ is compact and convex.
\end{assm}

\begin{assm}
  The regression function $g(x,\theta)$ is a Borel measurable real function on $R^p$ for each $\theta$ and is twice continuously differentiable with respect to $\theta$ for each $x$.
\end{assm}

\begin{assm}
  Let  $\|\cdot\|$ represent the Euclidean norm.
\begin{eqnarray*}
  && E\left\{\sup_{\theta\in \Theta} g^2(X,\theta)\right\}<\infty , \\
  && E\left\{\sup_{\theta\in\Theta} \left\| \frac{\partial g(X,\theta)}{\partial \theta}\frac{\partial g(X,\theta)}{\partial \theta^\t} \right\|\right\}<\infty,  \\
  && E\left\{\sup_{\theta\in\Theta} \left\| [Y-g(X,\theta)]^2 \frac{\partial g(X,\theta)}{\partial \theta}\frac{\partial g(X,\theta)}{\partial \theta^\t} \right\|\right\} < \infty, \\
  && E\left\{\sup_{\theta\in\Theta}\left\| [Y-g(X,\theta)]^2 \frac{\partial^2 g(X,\theta)}{\partial\theta\partial\theta^\t} \right\|\right\}<\infty.
\end{eqnarray*}
\end{assm}

\begin{assm}
  There exists a unique minimizer $\theta^\ast$ such that
\begin{eqnarray*}
  \theta^\ast = \arg\inf_{\theta\in \Theta} E\{Y-g(X,\theta)\}^2 .
\end{eqnarray*}
Under the null hypothesis, $\theta^\ast$ is an interior point of $\Theta$.
\end{assm}
\begin{assm}
  The matrix $E\{\dot g(X,\theta^\ast) \dot g(X,\theta^\ast)^\t\}$
is nonsingular.
\end{assm}

\begin{assm}
  The independent random vector $X$ satisfies the linearity condition: $E(X\cdn B^\t X) = P_B^\t(\Sigma) X$, where $B\in R^{p\times q}$ is any basis of $S_{\eta\cdn X}$.
\end{assm}

\begin{assm}
Let $\tilde m_1(t) = E(it \exp(it\v) X \ell )$, $\tilde m_2(t) = E(it X\dot g^\t \exp(it\v)) G^{-1} E(\dot g\ell)$ and $\tilde m(t) = \tilde m_1(t) - \tilde m_2(t)$. The matrix $\tilde M = \int \tilde m(t) \tilde m^H(t) d F_\v(t) $ is nonsingular.
\end{assm}

\begin{assm}
  The kernel function $K(\cdot)$ is a symmetric nonnegative and continuous function which is bounded and $\int K(u)du = 1$.
\end{assm}

\begin{assm}
  The bandwidth in kernel estimation satisfies $h\to 0$ and $nh^p\to \infty$ as $n\to 0$.
\end{assm}

Conditions $1-6$ are commonly required for the consistency and asymptotic normality of the least squares estimation for the parameter $\theta$, see \cite{zheng1996consistent}.
Condition $7$ is the linearity condition for the independent vector $X$ such that the target matrix $M$ is contained in the central subspace $S_{\eta\cdn X}$.
Condition $8$ is assumed for the target matrix such that the indicative dimension can be well identified  and then can be used for Theorems~\ref{thm: q under local} and  \ref{thm: asymptotic under local alternatives}.
Conditions $9$ and $10$ are commonly used conditions on the kernel estimation and the asymptotic normality of the corresponding test statistic. 

\bibliographystyle{Chicago}

\bibliography{ref_zerodim}

\end{document}